\documentclass[12pt,preprint]{aastex}

\newcommand{\degrees}{$^{\circ}$}

\newcommand{\um}{$\mu$m}

\slugcomment{This draft \today}
\shorttitle{COMPLETE Dust Emission Maps}
\shortauthors{Schnee et al.}

\begin{document}

\title{A COMPLETE Look at The Use of IRAS Emission Maps to Estimate 
       Extinction and Dust Temperature}
\author{Scott L. Schnee, Naomi A. Ridge, Alyssa A. Goodman \and Jason
        G. Li}
\affil{Harvard-Smithsonian Center for Astrophysics, 60 Garden St.,
       Cambridge, MA, 02138, USA}

\begin{abstract}
	We have created new dust temperature and column density maps
of Perseus, Ophiuchus and Serpens using 60 and 100 \um\ data from the
IRIS recalibration of IRAS data.  We describe an optimized method for
finding the dust temperature, emissivity spectral index, and optical
depth using optical and NIR extinction maps.  The creation of these
temperature and extinction maps (covering tens of square degrees of
molecular clouds) is one of the first results from the ongoing
COordinated Molecular Probe Line Extinction Thermal Emission
(COMPLETE) Survey of Star Forming regions.  However, while the
extinctions derived from the IRIS emission maps are globally accurate,
we warn that FIR emission is not a good proxy for extinction on the scale
of one pixel ($\sim$5\arcmin).

	In addition to describing the global dust properties of these
clouds, we have found two particularly interesting features in the
column density and temperature maps.  In the Ophiuchus dark cloud
complex, the new dust temperature map shows a little known warm
(25\,K) dust ring with 2\,pc diameter.  This shell is approximately
centered on the B star $\rho$-Ophiuchus, 1\degrees\ north of the
well-studied $\rho$-Oph star-forming cluster.  In Perseus, the column
density map shows a 10\,pc diameter ring, a feature not apparent in
the filamentary chain of clouds seen in molecular gas.  These rings
are further discussed in detail in companion papers to this one.
\end{abstract}

\keywords{ISM: clouds --- dust, extinction --- surveys}

\section{Introduction}
\label{INTRO}

	The goal of the COMPLETE Survey is to use a carefully chosen
set of observing techniques to fully sample the density, temperature
and velocity structure of three of the five large star-forming
complexes observed in the NASA-sponsored Spitzer Legacy Survey ``From
Molecular Cores to Planet-forming Disks'' (c2d) described in
\citet{Evans03}.

	The c2d Survey, started in late 2003, is producing
high-resolution infrared spectroscopy, and near- through far-infrared
images of each of its 5 pc scale target complexes. COMPLETE is
providing fully-sampled milimeter spectral-line, extinction, and
thermal emission maps for the same regions, at arcminute resolution or
better \citep{Goodman04}.  Phase II of COMPLETE, begun recently,
provides higher-resolution observations using the same suite of
techniques for a large subset of the high-density cores evident at
lower resolution.

	In this paper we present new temperature and extinction maps
of the Perseus, Serpens and Ophiuchus star-forming regions, produced
from 60 and 100 \um\ flux-density maps obtained from IRIS (Improved
Reprocessing of the {\it IRAS} Survey), and normalized using
optical/NIR extinction maps generated by \citet{Alves05} and
\citet{Cambresy99}.  Previous all-sky thermal dust emission maps by
\citet{Schlegel98} [SFD] include these regions.  However, the SFD maps
were based on low-resolution temperature data and optimized for low
extinction regions.  It has been established that the SFD maps lose
accuracy at $A_V > 0.5$\,mag \citep{Arce99b,Cambresy05}, so they are
not adequate for mapping column density in the high-extinction areas
targeted in the COMPLETE survey.  We show here that reanalysis of the
IRAS data can yield column density and color temperature maps that are
more accurate at high and low levels of extinction on the large scale,
though their use on the scale of the data ($\sim$ 5\arcmin) is
more limited.

	In Section \ref{DATA} we describe the IRIS flux measurements
and NIR/optical-based extinction maps.  In Sections \ref{EQUATIONS}
and \ref{CONSTANTS} we present and explain the equations used to
determine the the dust optical depth and conversion to visual
extinction, and the parameters needed in those equations.  In Section
\ref{RESULTS} we compare the results of our analysis technique to
those from other extinction tracers.  The results are summarized in
Section \ref{SUMMARY}.

\section{Data} 
\label{DATA}
\subsection{IRAS/IRIS}
\label{ISSA}

	IRIS\footnote{IRIS comprises a machine-readable atlas of the
sky in the four IRAS bands at 12, 25, 60, and 100 \um} images of
flux-density at 60 and 100 \um\ were obtained for each of our three
target regions from the IRIS website \citep{Miville05}.  Image details
are given in Table \ref{DEFTAB}. The maps have units of MJy sr$^{-1}$,
are made with gnomic projection and have spatial resolution smoothed
to 4.3\arcmin\ at 100 \um.

	IRIS data offer excellent correction for the effects of
zodiacal dust and striping in the images and also provide improved
gain and offset calibration.  Earlier releases of the IRAS data did
not have the appropriate zero point calibration, which can have
serious consequences on the derived dust temperature and column
density \citep{Arce99a}.  To test the zero point calibration of the
IRIS data, we allowed two free parameters in the fit (one each for the
60 and 100 micron zero points) to convert IRIS fluxes to visual
extinction (see Section \ref{CONSTANTS}).  We find that values of the
free parameters are consistent with zero, and thus provide independent
evidence that the zero point calibrations of the IRIS data at 60 and
100 \micron\ are correct.

\subsection{Optical and NIR Extinction Maps}
\label{OIR}

	Here we describe the NIR and optical extinction maps used to
recalibrate the IRIS maps presented in this paper.  We have obtained
extinction maps of Serpens and Ophiuchus from \citet{Cambresy99},
which are based on an optical star counting method with variable
resolution (but we regrid these maps to the constant resolution of the
IRIS images).  The optical photometry data used by Cambr{\' e}sy come
from the USNO-PMM catalogue \citep{Monet96}.  Extinction maps of
Perseus and Ophiuchus have been constructed from the 2MASS point
source catalog as part of COMPLETE by \citet{Alves05} using the
``NICER'' algorithm, which is a revised version of the NICE method
described in \citet{Lada94} and \citet{Lombardi01}.  NICE and NICER
combine direct measurements of near-infrared color excess and certain
techniques of star counting to derive mean extinctions and map the
dust column density distribution through a cloud \citep{Lombardi01}.
The 2MASS survey provides the NIR $J,H$ and $K_s$ colors of background
stars that have been reddened by the molecular cloud.  With these
measured colors, and knowledge of the intrinsic colors of these stars
(measured in a nearby, non-reddened contol field), the amount of
obscuring material along the line of sight to each star can be
determined.  In NICER, the extinction values are calculated for a
fixed resolution, which means uncertainties vary from pixel to pixel.
Maps of uncertainty for the regions considered are shown in
\citet{Alves05}.

	In the Ophiuchus molecular cloud, we have both 2MASS/NICER and
optical star counting based extinction maps.  It is important to note
that the two methods {\it do not} give equivalent extinctions.  A plot
of the optical vs NIR extinctions is shown in Figure
\ref{OPHCAM2MASS}.  A linear fit to the data shows that the slope of
the line relating the two quantities is very close to unity (as
expected), {\it but} there is an offset of roughly 0.71 magnitudes of
visual extinction, with the 2MASS derived data points being
systematically higher than the optically derived extinctions.  The
1$\sigma$ scatter on the least squares fit between the two methods is
0.7 magnitudes in $A_V$.  Because these methods produce such different
results, care must be taken when using any one extinction map as a
model to determine dust properties as described in Section
\ref{EQUATIONS}.  In Ophiuchus, where we have both optical and NIR
extinction maps, we run the global fitting algorithm for each data set
separately, and report both sets of results in Table \ref{FITTAB}.

	In Perseus we only have a NIR based extinction map, and in
Serpens we only have an optically derived extinction map, so no
intercomparison of the calibration methods is possible for these
clouds.  We expect that the 2MASS derived extinctions are more
accurate than the star counting method of Cambr{\' e}sy, although in
both cases the zero point calibration can be difficult to assess.  The
optical star counting method chooses the average minimum value in the
map as the ``zero level'' of extinction \citep{Cambresy99}, and the
NICER method uses the average minimum H-K color to determine the
minimum extinction in the map.  Both methods rely on the minimum in
their map actually corresponding to zero extinction, which if untrue,
will cause both methods to underestimate the true extinction.  In
Ophiuchus, because the NICER algorithm gives higher extinctions, it is
likely that the optical star counting algorithm has underestimated the
true extinction by approximately 0.71 magnitudes.  Our decision to
trust the NIR based extinctions over the star counting extinctions is
supported by discussion with Drs. Cambr{\' e}sy and Alves (private
communication).  To account for the 0.71 mag offset, we have added
0.71 magnitudes to the extinction values derived by \citet{Cambresy99}
in our optically calibrated calculations for Ophiuchus in Sections
\ref{EQUATIONS} and \ref{CONSTANTS}.  

	We expect the Serpens extinction map to suffer from the same
``zero level'' problem as the Ophiuchus optical extinction map, and
therefore include a free parameter in the fit between the optical
extinction map and the IRIS emission maps to account for this
possibility.  We find the best fit occurs between the emission based
extinction map and the optical star count based extinction map when
the latter map has a constant value of 2.4 magnitudes of visual
extinction added to each pixel.

\section{Basic Formulae}
\label{EQUATIONS}

	The basic method that we use to calculate the dust color
temperature and column density from the IRAS 60 and 100 \um\ flux
densities is similar to \citet{Wood94} and \citet{Arce99a}.  The
temperature is determined by the ratio of the 60 and 100 \um\ flux
densities.  The column density of dust can be derived from either
measured flux and the derived color temperature of the dust.  The
calculation of temperature and column density depend on the values of
three parameters: two constants that determine the emissivity spectral
index, and the conversion from 100 \um\ optical depth to visual
extinction.  We are able to solve for these parameters explicitly
because we have an independent estimate of visual extinction, as will
be explained in Section \ref{CONSTANTS}.

	The dust temperature $T_d$ in each pixel of a FIR image can be
obtained by assuming that the dust in a single beam is isothermal and
that the observed ratio of 60 to 100 \um\ emission is due to blackbody
radiation from dust grains at $T_d$, modified by a power-law
emissivity spectral index.  The flux density of emission at a
wavelength $\lambda_i$, is given by
\begin{equation}
\label{FLUXEQN}
F_i=
[\frac{2hc}{\lambda_i^3(e^{hc/(\lambda_ikT_d)}-1)}]
N_d\alpha\lambda_i^{-\beta}\Omega_i,
\end{equation}
where $N_d$ represents column density of dust grains, $\alpha$ is a
constant that relates the flux to the optical depth of the dust,
$\beta$ is the emissivity spectral index, and $\Omega_i$ is the solid
angle subtended at $\lambda_i$ by the detector.

	Following \citet{Dupac03}, we use the equation,
\begin{equation}
\label{BETAEQN}
\beta = \frac{1}{\delta + \omega T_d}
\end{equation}
to describe the observed inverse relationship between temperature and the
emissivity spectral index.  The parameters ($\delta$ and $\omega$) are
derived separately for each cloud and subregion considered in this
paper. 

	With the assumptions that the dust emission is optically thin
at 60 and 100 \um\ and that $\Omega_{60} \simeq \Omega_{100}$ (true 
for IRIS images), we can write the ratio, $R$, of the flux densities 
at 60 and 100 \um\ as
\begin{equation}
\label{TEMPEQN}
R=0.6^{-(3+\beta)}\frac{e^{144/T_d}-1}{e^{240/T_d}-1}.
\end{equation}

Once the appropriate value of $\beta$ is known, one can
use Equation \ref{TEMPEQN} to derive $T_d$.  The value of $\beta$
depends on such dust grain properties as composition, size, and
compactness.  For reference, a pure blackbody would
have $\beta = 0$, amorphous layerlattice matter has $\beta \sim 1$, and 
metals and crystalline dielectrics have $\beta \sim 2$.

	Once the dust temperature has been determined, the 100 \um\
dust optical depth can be calculated as follows:
\begin{equation}
\label{TAUEQN}
\tau_{100}=\frac{F_\lambda(100 \mu m)}{B_\lambda(100 \mu m, T_d)},
\end{equation}
where $B_\lambda$ is the Planck function and $F_\lambda(100 \mu m)$ is
the 100 \um\ flux.

	The 100 \um\ optical depth can then be converted to V-band
extinction using: 
\begin{equation}
\label{AVEQN}
A_V=X \tau_{100}
\end{equation}
where $X$ is a parameter relating the thermal emission
properties of dust to its optical absorption qualities.  

\section{Derivation Of Constants}
\label{CONSTANTS}

	We can derive the values of the three parameters ($\delta$,
$\omega$ and $X$) from the IRIS emission maps and extinction maps from
\citet{Cambresy99} and \citet{Alves05}.  Each optical and NIR
extinction map described in Section \ref{OIR} is used as a ``model''
extinction map to fit the IRIS-implied column density map using the
three adjustable parameters which are explained in Section
\ref{EQUATIONS}.  The IDL task AMOEBA was used to simulateously fit
all three parameters with the downhill simplex method of
\citet{Nelder65}.  Each combination of these three parameters is used
to determine the dust temperature and column density at each point in
the map (using the formulae in Section \ref{EQUATIONS}).  The
parameter values determined by this method are those that create a
FIR-based extinction map that best matches the NIR color excess or
optical star count derived extinction map.  This is a statistical
point-by-point match, not a spatial match to features in the
NIR/optical extinction map.  As explained in Section \ref{OIR}, we
solve for a fourth parameter in Serpens, which is the zero level of
the optical extinction map.

	Each cloud is considered separately, so the values of
$\delta$, $\omega$ and $X$ are different for each cloud.
Because Ophiuchus has two independent extinction maps, one from
optical star counting (adjusted by the 0.71 mag offset) and one from
the NIR color excess method, it has two sets of constants derived for
it.  In this paper we assume that the values of the three parameters
are constant within each image, though of course this does not have to
be the case.  For instance, it may be the case that areas of
especially high or low column density do not share the same visual
extinction conversion factor, $X$.  Our values for the three
parameters for each region fit (four in Serpens) are presented in
Table \ref{FITTAB}.

\section{Results and Discussion}
\label{RESULTS}

\subsection{Assumption of Thermal Equilibrium}
\label{EQUIL}

	The derivations of dust temperature and column density from
IRIS data rely on the assumption that the dust along each line of
sight is in thermal equilibrium.  However, the dust models of
\citet{Desert90} show that the emission at 60 and 100 \micron\ have
contributions by big dust grains (BGs) and very small dust grains
(VSGs).  The VSGs are not in thermal equilibrium, and emit mostly in
the 60 micron band, while the BGs are the primary contributors at 100
microns and longer.  Because of the emission from the VSGs,
determining the temperature of the dust from the ratio of 60 and 100
\micron\ fluxes can yield temperatures that are systematically high.
An empirical method for determining the color temperature and optical
depth of dust using the 60 and 100 micron bands from IRAS is presented
in \citet{Nagata02}, but as this method has been calibrated for
galaxies and not molecular clouds, we do not employ their method.

	In order to remove the contribution of VSGs to the 60 \micron\
flux, we calibrate our temperature maps to those derived by
\citet{Schlegel98} [SFD], which are not currupted by VSG emission
because they used maps at 100 \micron\ and longer to derive their
temperatures.  We smooth the IRIS images to the resolution of the SFD
temperature maps, and calculate the temperature based on 100\% of the
100 \micron\ flux and a lesser percentage of the 60 \micron\ flux,
assuming that $\beta = 2$ (as assumed by SFD).  The fraction of the 60
\micron\ flux was chosen so as to best match our derived temperatures
with the SFD temperatures.  The remaining 60 \micron\ flux comes from
the VSGs.  Note that the temperatures we use in our calculations of
column density do not assume that $\beta = 2$; we use the temperature
dependent form of $\beta$ described in Section \ref{EQUATIONS}.
However, the 60 \micron\ flux used in Section \ref{EQUATIONS} is
adjusted by this determination.  By comparison to the SFD temperature
maps, we find that the average VSG contribution to the 60 \micron\
flux is 74\%, 72\% and 85\% in Perseus, Ophiuchus and Serpens,
respectively.

\subsection{Assumption of Uniform Dust Properties}
\label{ASSUME}

	The equations in this paper rely on the assumption that the
dust along each line of sight is characterized by a single
temperature, emissivity spectral index and emissivity.  This is a
simplification that our method requires, though we recognize that real
molecular clouds are much more complicated.
	
	In their FIR study of interstellar cold dust in the galaxy,
\citet{Lagache98} have found that there are at least two temperature
components to the dust population.  They find that the warmer
component, associated with diffuse dust, has a temperature around 17.5
K, while the colder component, associated with dense regions in the
ISM, has a temperature around 15 K.  The molecular clouds we study
here are expected to have a range of temperatures along some lines of
sight that is even wider than seen in \citet{Lagache98} because the
FIRAS data used in this study has a beam size of 7\degrees, which is
larger than the size of the maps for each cloud studied in this paper.
The color temperature that we derive from our isothermal assumption is
therefore biased, as is the optical depth.  This problem is especially
relevant in Serpens, which is only a few degrees above the galactic
center, so there are certainly multiple environments integrated into
each IRAS beam.  A method for determining the amount of 100 \micron\
flux associated with the cold component of the BGs in a molecular
cloud is explained in \citet{Abergel94}, \citet{Boulanger98} and
\citet{Laureijs91}.  Nevertheless, \citet{Jarrett89} find in Ophiuchus
that there is a very tight linear correlation between FIR optical
depth (determined in much the same way as presented here) and visual
extinction for $A_V \le 5$, so we trust that the errors introduced by
our method are not prohibitively large.

\subsection{Temperature and Column Density Maps}
\label{PUBLIC}

\begin{figure}
\epsscale{0.75}
\plotone{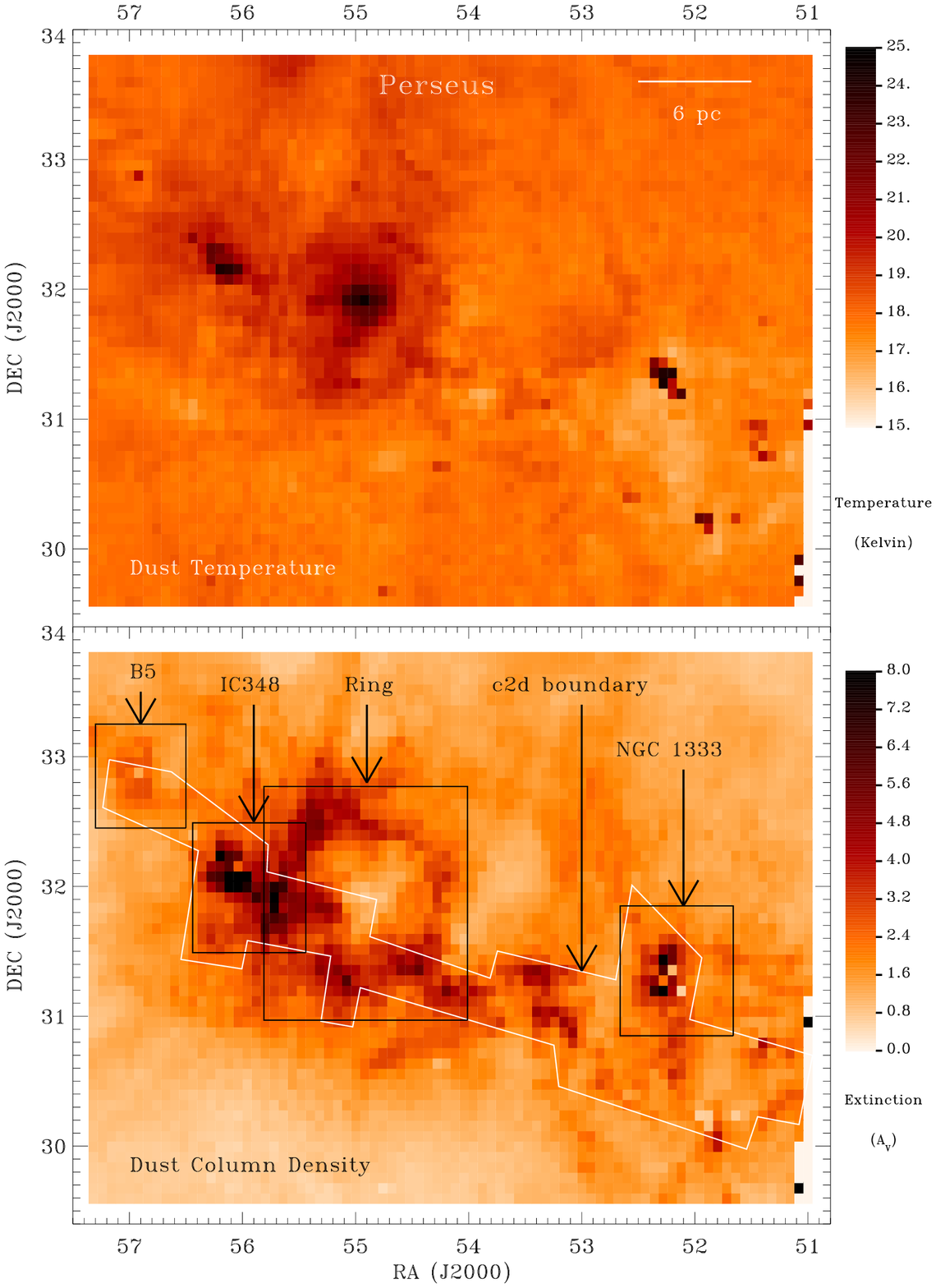}
\caption{Dust color temperature map (top) and column density map 
	(bottom) of Perseus created from IRIS 60 and 100 \um\
	images.  \label{PERTEMPEXT}}
\end{figure}

\begin{figure}
\epsscale{0.75}
\plotone{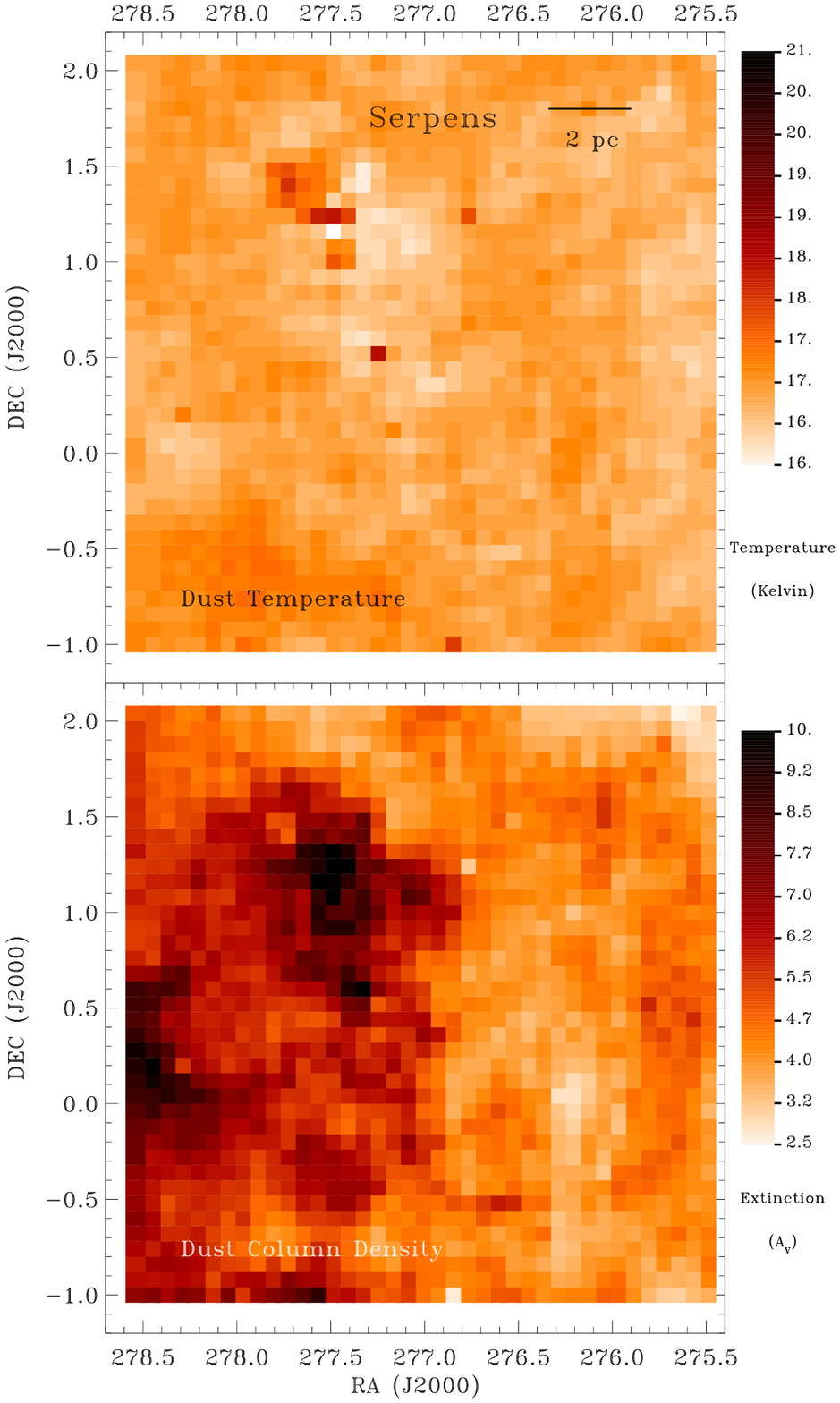}
\caption{Dust color temperature map (top) and column density map (bottom)
	 of Serpens created from IRIS 60 and 100 \um\ images.
	 \label{SERTEMPEXT}}
\end{figure}

\begin{figure}
\epsscale{0.75}
\plotone{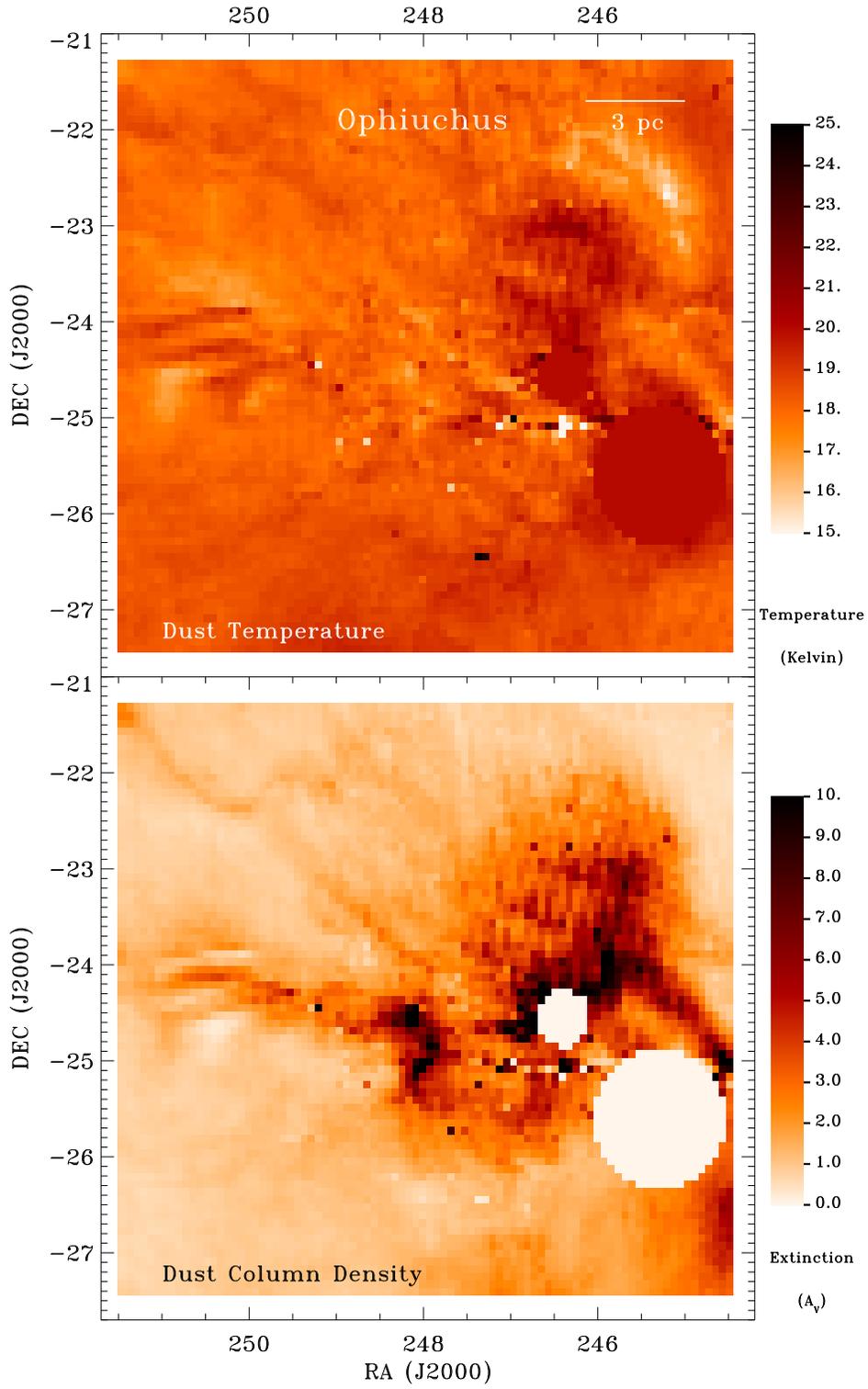}
\caption{Dust color temperature map (top) and column density map (bottom) 
	 of Ophiuchus created from IRIS 60 and 100 \um\ images.  The two
	 white circles have been excluded due to contamination from B stars.
	 \label{OPHTEMPEXT}}
\end{figure}

	The dust color temperature and column density maps derived
from our parameter fits and IRIS data are among the first publicly
available data products distributed by the COMPLETE team.  Temperature
and extinction maps of Perseus, Ophiuchus and Serpens derived from the
IRIS data are shown in Figures \ref{PERTEMPEXT}, \ref{SERTEMPEXT} and
\ref{OPHTEMPEXT}.  FITS files of these maps can be downloaded from the
COMPLETE web page\footnote{http://cfa-www.harvard.edu/COMPLETE}.

	In Perseus, there is a striking ring of emission that is
centered on a region of warm material.  This ring has been discussed
by \citet{Andersson00}, \citet{Pauls89} and \citet{Fiedler94} and is
discussed in more detail in a companion paper to this one
\citep{Ridge05}.  The Perseus ring does not stand out in the NICER
extinction map as visibly as in the IRIS column density map because
it is not a true column density feature.  It is difficult to determine
from emission maps alone if individual features are the result of
changing dust properties or column density enhancements.

	A warm dust ring is evident in Ophiuchus in the temperature
map, centered at the position 16:25:35, -23:26:50 (J2000).  This ring
was reported by \citet{Bernard93} in a discussion of the far-infrared
emission from Ophiuchus and Chameleon, but they did not investigate
its nature or possible progenitor.  The B star $\rho$-Ophiuchus and a
number of X-ray sources are projected to lie within this shell of
heated gas.  The Ophiuchus ring will be further discussed in a future
paper \citep{Li05}.

\subsection{Temperature and Extinction Distributions}
\label{TEMPEXT}

\begin{figure}
\epsscale{0.8}
\plotone{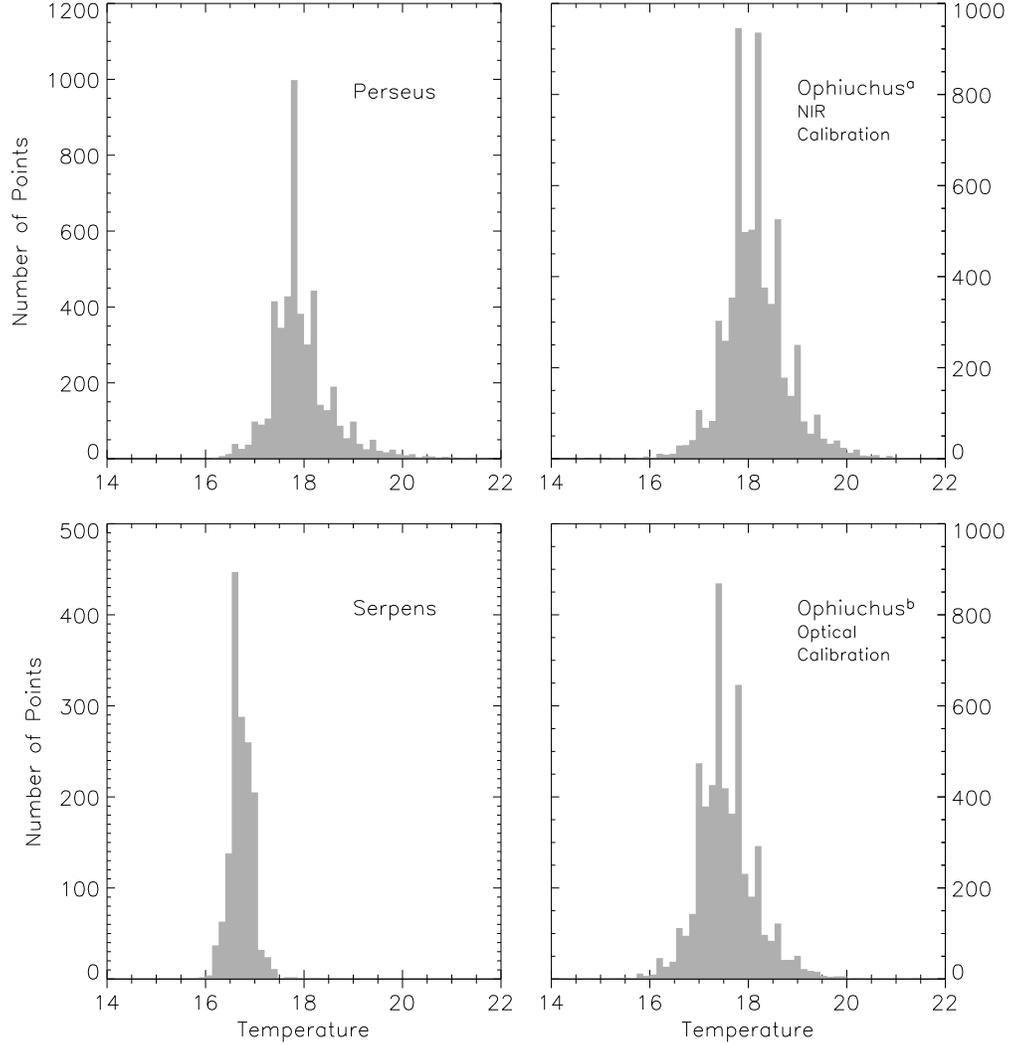}
\caption{The filled histograms show the temperature derived from the 
	60/100$\mu$m flux density ratio in Perseus, Serpens and Ophiuchus 
	after the calibration described in Section \ref{DATA}.  
	The Ophiuchus$^a$ histogram (top right) is calibrated to the 
	2MASS extinction map.  The Ophiuchus$^b$ histogram (bottom 
	right) is calibrated to the Cambr{\' e}sy data.
	\label{TEMPHIST}}
\end{figure}

	In Figure \ref{TEMPHIST} we show histograms of the dust
temperature in Perseus, Serpens, and Ophiuchus.  Each distribution
peaks near 17 K, and all except Serpens have a spread of several
degrees.  The dust temperatures that we derive are, to an extent,
calibrated with those derived by \citet{Schlegel98} (see Section
\ref{EQUIL}).

	It has been shown for Taurus by \citet{Arce99a} that using
IRAS 60 and 100 \um\ flux densities to determine dust column density
gives results consistent with other methods, such as the color excess
method (e.g. NICE/NICER), star counts (e.g. \citet{Cambresy99}), and
using an optical (V and R) version of the average color excess method
used by \citet{Lada94}.  Here we compare the IRIS derived extinction
maps of Perseus, Serpens, and Ophiuchus with maps created by some of
these other methods.

\begin{figure}
\epsscale{0.8}
\plotone{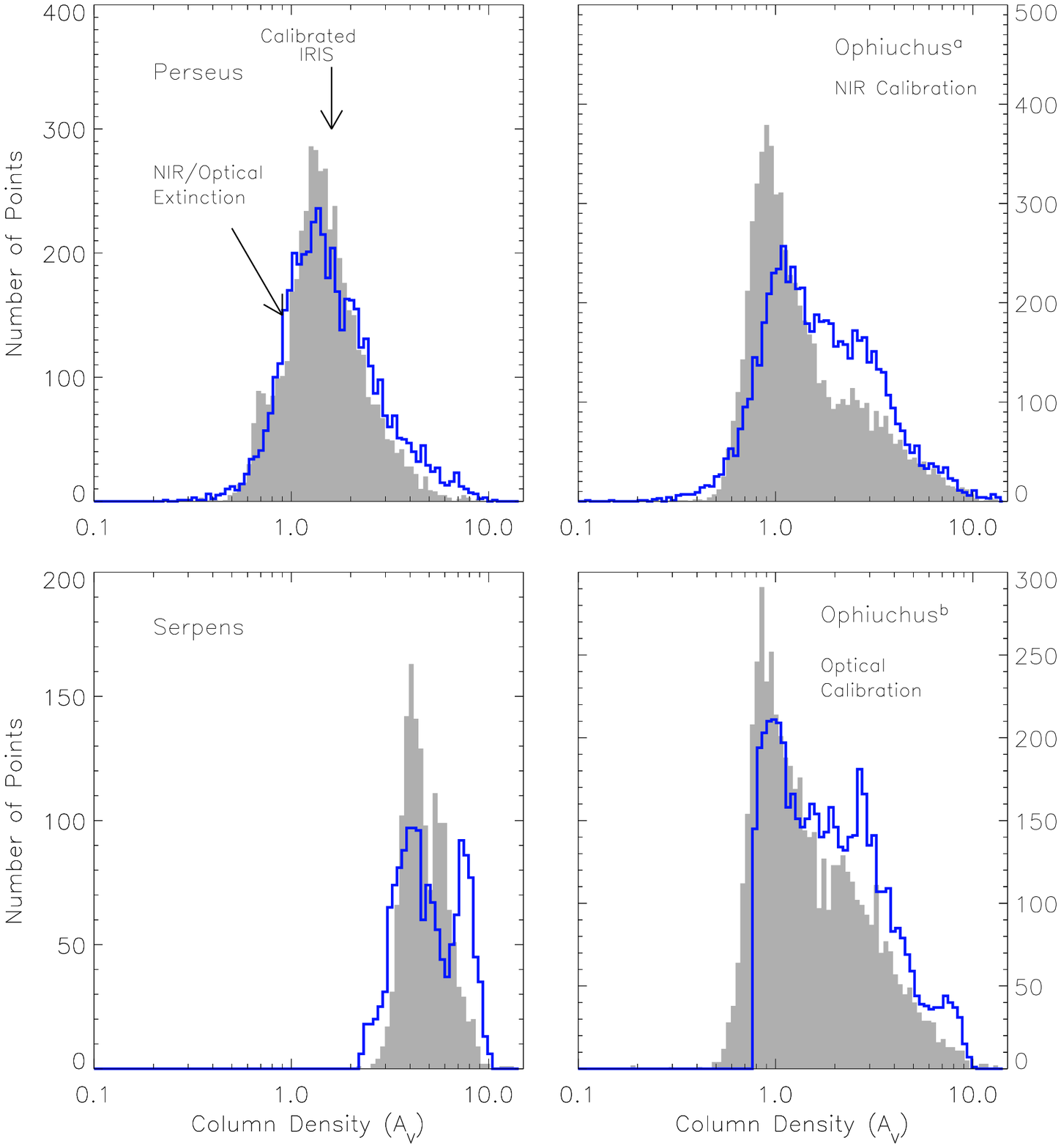}
\caption{The filled histograms show the extinction derived from the 
	60/100$\mu$m flux density ratio in Perseus, Serpens and Ophiuchus 
	after the calibration described in Section \ref{DATA}.  The blue 
	open histogram shows the NIR/optical derived extinction (from 
	\citet{Alves05} and \citet{Cambresy99}).  The Ophiuchus$^a$ 
	histogram (top right) is calibrated to the 2MASS extinction map.  
	The Ophiuchus$^b$ histogram (bottom right) is calibrated to the 
 	Cambr{\' e}sy data.
	\label{EXTHIST}}
\end{figure}

\begin{figure}
\epsscale{0.8}
\plotone{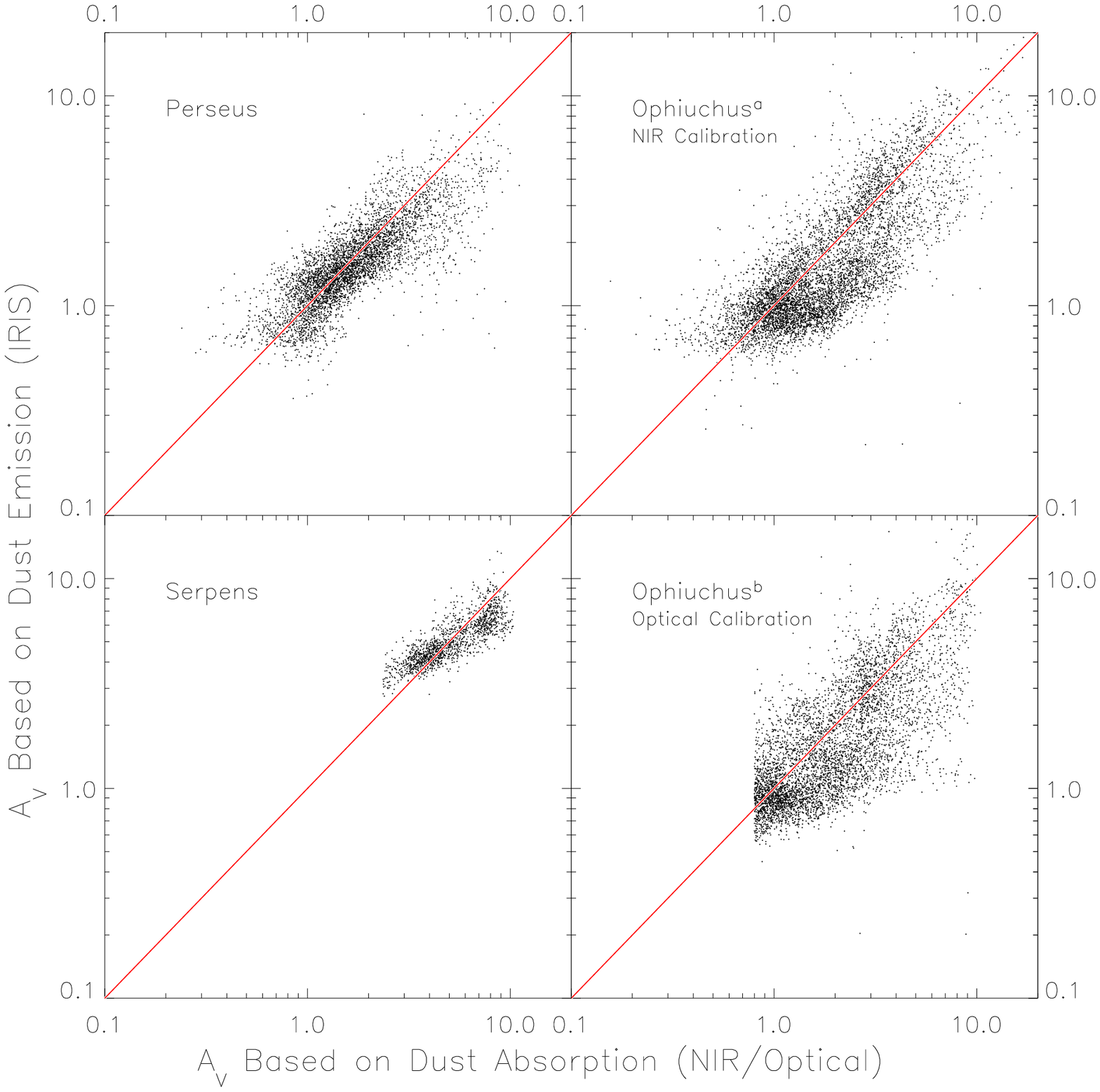}
\caption{Scatter plot of the corrected IRIS extinction (derived from the 
	60/100$\mu$m flux density ratio in Perseus, Serpens and Ophiuchus) 
	and the model extinction (from \citet{Alves05} and 
	\citet{Cambresy99}).  The Ophiuchus$^a$ IRIS extinction (top right)
	is calibrated to the 2MASS extinction map.  The Ophiuchus$^b$ 
	IRIS extinction (bottom right) is recalibrated to the 
	Cambr{\' e}sy data.  Note that the data are shown here in log-log 
	plots for clarity of display, but all the fitting is of linear (not 
	log) quantities.  The sharp cutoff in the Ophiuchus$^b$ panel is a
	result of having added 0.71 magnitudes to the optically derived
	extinction.  The sharp cutoff in the Serpens panel is a
	result of having added 2.2 magnitudes to the optically derived
	extinction.
	\label{EXTSCATTER}}
\end{figure}

	Our method requires the FIR-derived extinction to best match
the NIR or optical-derived extinction, and the global agreement can be
seen in Figure \ref{EXTHIST}.  However, the point-to-point extinction
values can be significantly different between the two methods, as
shown in Figure \ref{EXTSCATTER}.  The large scatter in these
extinction plots ($1\sigma \simeq 1$ mag $A_V$) is likely the result
of the various assumptions used in our calculations.  For instance, it
is unlikely to be the case that all of the dust along a given line of
sight can be well characterized by a single temperature or emissivity
spectral index, especially along lines of sight through the denser
regions of the molecular clouds.  It is also likely that the dust
optical depth to visual extinction conversion factor ($X$) is
not constant throughout a cloud volume.  The flux detected by IRAS
comes preferentially from warmer dust, while the extinction maps made
from NIR and optical data have no temperature bias, so it is also
possible (and in many regions likely) that the dust doing most of the
FIR emitting is not the same dust responsible for most of the
extinguishing at shorter wavelengths.  The scatter for each cloud is
shown in Table \ref{SCATTERTAB}.

	As a test to see if the parameters determined for one cloud
can be successfully used to describe another, we used the Perseus
fits for $\delta$, $\omega$ and $X$ for the Ophiuchus 60 and 100
\um\ IRIS maps and compared the derived extinction to the 2MASS
extinction map of Ophiuchus.  The median point-to-point difference was
0.8 magnitudes of visual extinction, with a standard deviation of 1.9
magnitudes.  When the optimized parameters for Ophiuchus are used, the
median point-to-point difference is only 0.2 mag, with a scatter of
1.2 mag.  We conclude that {\it the fits for one cloud are unlikely to
be appropriate for other clouds}, and therefore caution should be used
in attempts to estimate extinction within molecular clouds from IRAS
emission maps alone.

  	We have derived separate values for $\delta$, $\omega$ and $X$
for various regions within Perseus to see if the dust properties are
significantly different there than in the cloud as a whole.  The
regions considered were B5, IC348, NGC1333, the emission ring, and the
area surveyed by the c2d Spitzer Legacy project.  The locations and
sizes of the Perseus sub-regions are shown in Table \ref{DEFTAB}.  The
values of $\delta$, $\omega$ and $X$ for these sub-regions are shown
in Table \ref{FITTAB} and displayed in Figure \ref{PERTEMPEXT}.  The
value of $\delta$ varies by about 10 percent, $\omega$ varies by about
20 percent and $X$ varies by 20 percent in these sub-regions of
Perseus.  NGC1333 varies much more significantly from the other
sub-regions of Perseus.

	We conclude that the method described here for converting dust
emission to visual extinction can be used with confidence to find
regions with high or low extinction and to determine the average
extinction in large areas ($\sim$0.25 square degrees).  However, the
extinction determined in any individual 5\arcmin\ pixel {\it should
not be trusted} to represent the true absolute extinction to better
than $\sim$ 2.5 magnitudes (2$\sigma$) of visual extinction.  This is
made clear by the significant variability in the emissivity spectral
index of dust and the conversion from dust optical depth to visual
extinction between clouds and within clouds, as shown in Tables
\ref{FITTAB} and \ref{SCATTERTAB}.

\subsection{Emissivity Spectral Index}
\label{EMISSPEC}

\begin{figure}
\epsscale{1.0}
\plotone{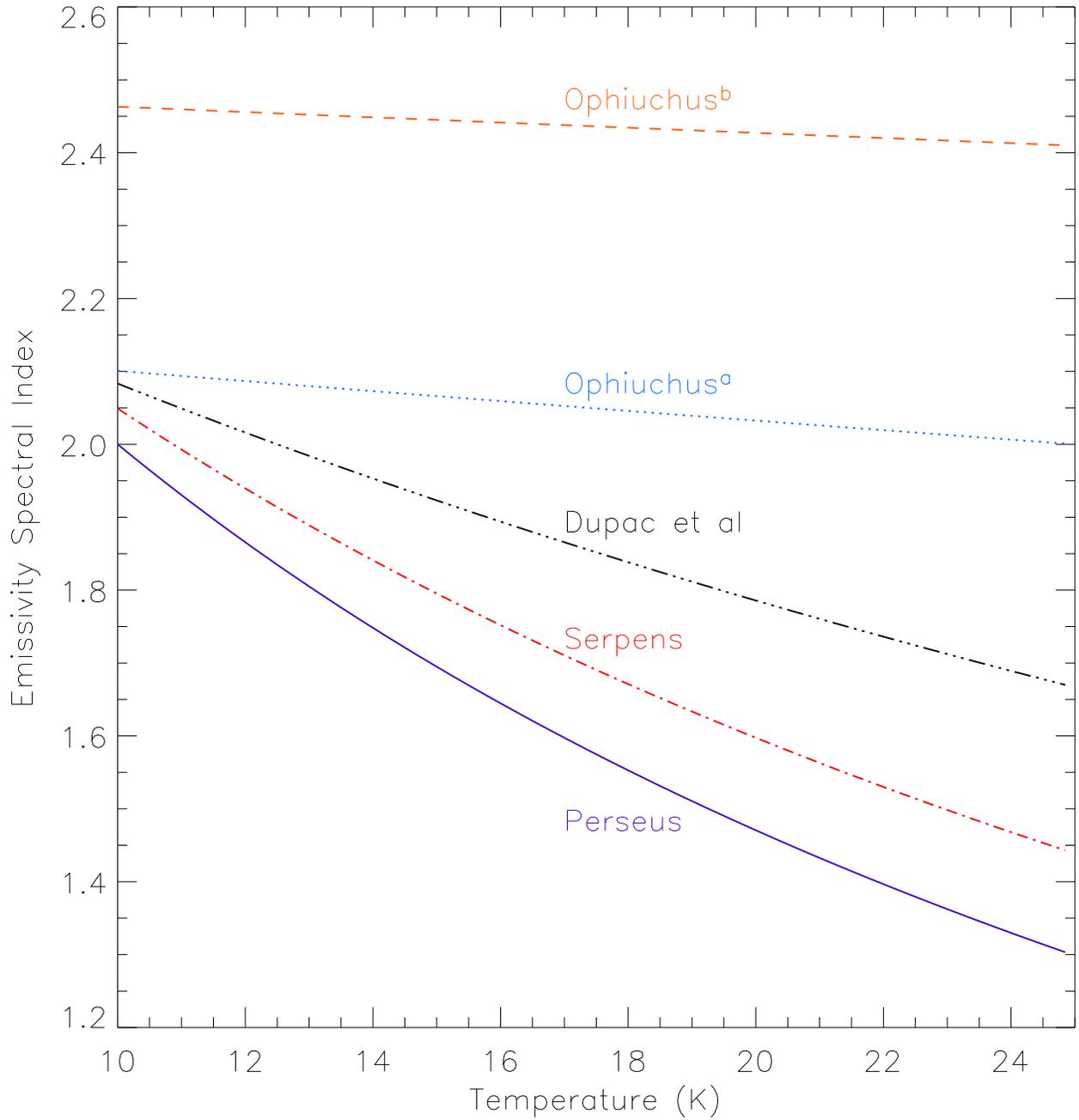}
\caption{Plot of the emissivity spectral index versus temperature for
         Perseus, Ophiuchus and Serpens.  The Ophiuchus$^a$ curve is
         for the IRIS data calibrated to the 2MASS extinction map.  
	 The Ophiuchus$^b$ curve is calibrated to the Cambr{\' e}sy data.
         The Dupac et al curve is the best fit curve to a collection of 
	 FIR/sub-mm emission regions as presented in \citet{Dupac03}.
         \label{BETAPLOT}}
\end{figure}

	The emissivity spectral index that we use in this paper varies
with temperature as shown in Equation \ref{BETAEQN}.  The values that
we find for the parameters $\delta$ and $\omega$ are shown in Table
\ref{FITTAB}.  Their values as determined by \citet{Dupac03} are 0.4
and 0.008 respectively, which comes from a much broader range of
environments and many more measurements of the FIR/sub-mm flux.
Figure \ref{BETAPLOT} shows the emissivity spectral index that we find
for each cloud plotted along with the curve from \citet{Dupac03}.  The
emissivity spectral index is considerably larger in Ophiuchus than in
Serpens, which is somewhat higher than $\beta$ in Perseus.  The
best-fit curve to the wider range of environments in \citet{Dupac03}
falls between our Serpens and Ophiuchus curves.

\section{Summary}
\label{SUMMARY}

	We have described a new method that uses NIR color excesses or
optical star counts to constrain the conversion of IRIS 60 and 100
\um\ data into color temperature and column density maps.  Our method
also derives the dust emissivity spectral index and the conversion
from dust 100 \micron\ optical depth to visual extinction.  We test
the IRIS 60 and 100 \micron\ zero points and confirm that, unlike
earlier releases of IRAS data, the IRIS recalibration is properly zero
point corrected.  We find the the very small grain contribution to the
60 \micron\ flux is significant.

	The dust temperature maps of Perseus, Serpens and Ophiuchus
available through the COMPLETE website should be the best dust
temperature maps of large star forming regions at 5\arcmin\ resolution
created to date.  The dust temperatures derived here are dependent on
the emissivity spectral index of the dust, which is solved globally
for each molecular cloud, and recorded in Table \ref{FITTAB}.

\begin{figure}
\plotone{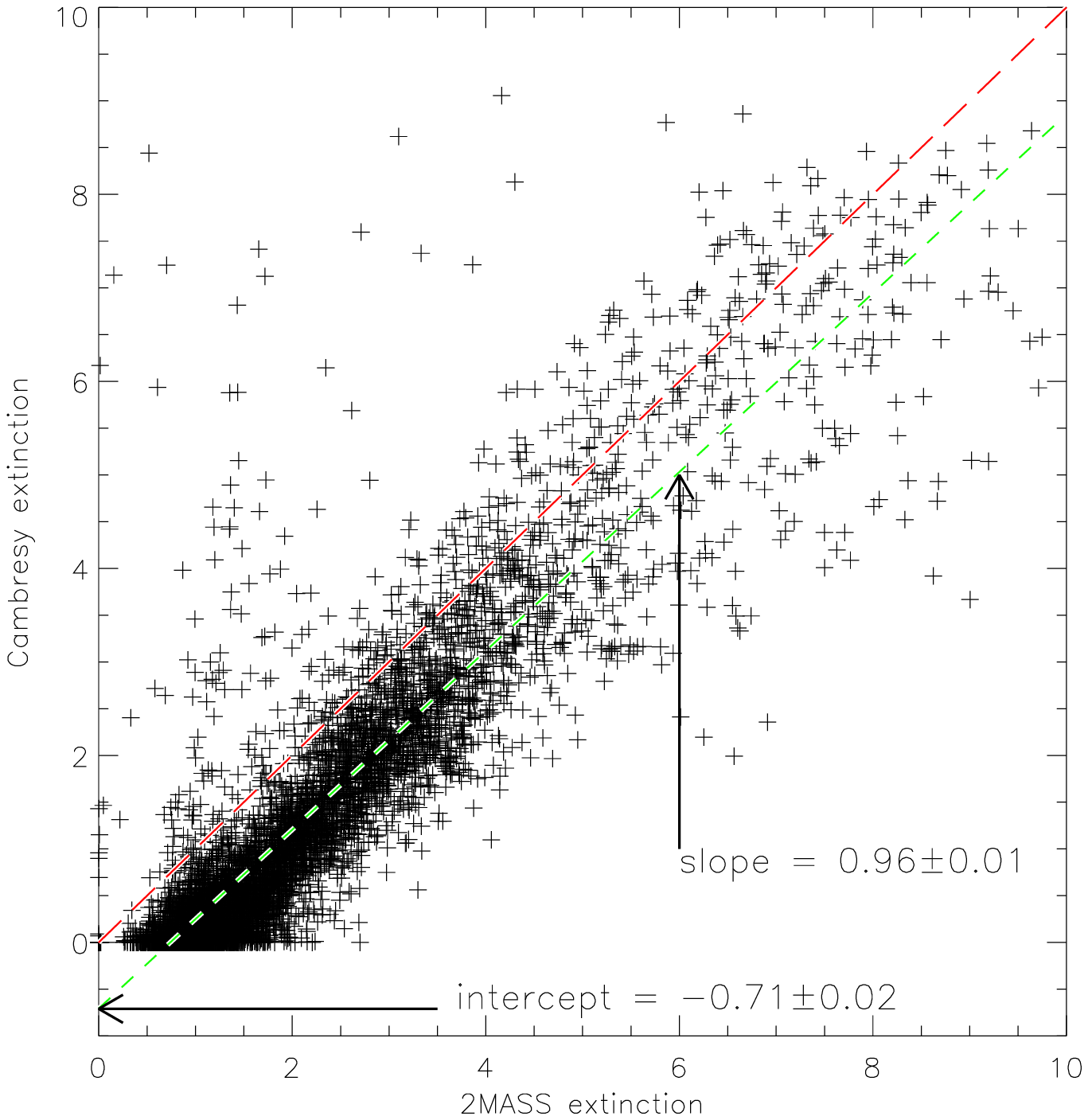}
\caption{Scatter plot of the Ophiuchus optical star counting 
         \citep{Cambresy99} and NIR color excess method 
	 \citep{Alves05} extinctions in Ophiuchus.  The green line
         (short dashes) is the best fit line to the data sets, with the 
	 equation shown at bottom of the image.  The red (1:1) line 
	 (long dashes) is there only to guide the eye.  Nearly 60 percent 
	 of the data points have 2MASS extinctions less than 2.
         \label{OPHCAM2MASS}}
\end{figure}

	Our work here indicates that one can not confidently convert
dust 100 \micron\ optical depth to visual extinction without the
benefit of having an extinction map made in an alternative manner to
use as a model because the conversion constant $X$ varies
significantly from cloud to cloud, and even within a cloud (see Table
\ref{FITTAB}).  Using the values for the dust emissivity spectral
index and visual extinction conversion derived for one cloud results
in a significant miscalculation of the extinction in other clouds.
Even with the NICER and optical star counting extinction maps used
here to calibrate the IRAS data, there is significant point-to-point
variation between the two estimates (see Figure \ref{OPHCAM2MASS}).
In fact, we have shown that in Ophiuchus the optical star count method
employed by \citet{Cambresy99} has an offset from the NICER method
applied to the 2MASS data set of $\sim$ 0.71 magnitudes in A$_V$, with
a scatter of 0.7 magnitudes, so the issue is not simply one of dealing
with the FIR dust properties in molecular clouds.

	In addition, we have identified two ring structures, one in
Perseus and one in Ophiuchus, which are discussed in upcoming papers
\citep{Ridge05} and \citep{Li05} respectively.  

\acknowledgments 
	We would like to thank the referee Laurent Cambr{\' e}sy for
many useful comments.  This research has made use of the NASA/ IPAC
Infrared Science Archive, which is operated by the Jet Propulsion
Laboratory, California Institute of Technology, under contract with
the National Aeronautics and Space Administration.  This publication
makes use of data products from the Two Micron All Sky Survey, which
is a joint project of the University of Massachusetts and the Infrared
Processing and Analysis Center/California Institute of Technology,
funded by the National Aeronautics and Space Administration and the
National Science Foundation.  This material is based upon work
supported under a National Science Foundation Graduate Research
Fellowship.

\bibliographystyle{apj}
\bibliography{refs}

\clearpage


\clearpage

\clearpage

\clearpage

\clearpage

\clearpage 
\begin{deluxetable}{cccc}
\tabletypesize{\scriptsize}
\tablecaption{IRIS Image Details \label{DEFTAB}}
\tablewidth{0pt}
\tablehead{
\colhead{Region}     &
\colhead{Center RA}  & 
\colhead{Center Dec} & 
\colhead{Image Size} \\
\colhead{}           &
\colhead{(degrees)}  &
\colhead{(degrees)}  &
\colhead{(degrees)}}
\startdata
Perseus      &  54.30 &  31.85 & 6.6$\times$4.5 \\
Perseus Ring &  54.91 &  31.87 & 1.8$\times$1.8 \\
B5           &  56.90 &  32.85 & 0.8$\times$0.8 \\
IC348        &  55.94 &  31.99 & 1.0$\times$1.0 \\
NGC1333      &  52.16 &  31.35 & 1.0$\times$1.0 \\
Ophiuchus    & 247.95 & -24.00 & 6.9$\times$6.8 \\
Serpens      & 277.00 &   0.50 & 3.0$\times$3.0 \\
\enddata
\tablecomments{All positions are given in equatorial J2000 equinox 
               coordinates.}
\end{deluxetable}

\clearpage
\begin{deluxetable}{ccccc}
\tabletypesize{\scriptsize}
\tablecaption{Best Fit IRAS/dust Parameters \label{FITTAB}}
\tablewidth{0pt}
\tablehead{
\colhead{Cloud}              & 
\colhead{$\delta$}           &
\colhead{$\omega$}           & 
\colhead{$X$}                &
\colhead{$A_V$ offset}       \\ 
\colhead{}                   &
\colhead{}                   &
\colhead{}                   &
\colhead{}                   &
\colhead{(mag $A_V$)}}
\startdata 
Perseus      & 0.32 & 0.0180 & 1100 & --  \\
B5           & 0.38 & 0.0140 & 1200 & --  \\
IC348        & 0.33 & 0.0160 & 1000 & --  \\
NGC1333      & 0.64 & 0.0010 & 1300 & --  \\
c2d Area     & 0.36 & 0.0140 & 1200 & --  \\
Ring         & 0.38 & 0.0150 &  990 & --  \\
Ophiuchus\tablenotemark{a} & 0.46 & 0.0016 & 510 & --   \\
Ophiuchus\tablenotemark{b} & 0.40 & 0.0006 & 350 & 0.71 \\
Serpens      & 0.35 & 0.0139 & 800 & 2.36 \\
\enddata
\tablecomments{Best fit values for the parameters used to convert IRIS
	       flux to visual extinction}
\tablenotetext{a}{IRAS flux normalized to the 2MASS/NICER extinction}
\tablenotetext{b}{IRAS flux normalized to the Cambr{\'e}sy extinction}
\end{deluxetable}

\clearpage
\begin{deluxetable}{ccccc}
\tabletypesize{\scriptsize}
\tablecaption{Scatter between IRAS implied extinction and 
              NIR/optical derived extinction \label{SCATTERTAB}}
\tablewidth{0pt}
\tablehead{
\colhead{Cloud}                             & 
\colhead{Median Difference}                 &
\colhead{1$\sigma$ Scatter Between Methods} &
\colhead{Scatter Between Methods}           &
\colhead{$\tau_{100}$/$A_V$}                \\
\colhead{}                                  &
\colhead{(mag $A_V$)}                       &
\colhead{(mag $A_V$)}                       &
\colhead{Percent}                           &
\colhead{10$^{-4}$ mag$^{-1}$}} 
\startdata
Perseus\tablenotemark{a}   & -0.1 & 0.9 & 30 &  9 $\pm$ 3  \\
Ophiuchus\tablenotemark{a} & -0.2 & 1.2 & 50 & 17 $\pm$ 9  \\
Ophiuchus\tablenotemark{b} & -0.2 & 1.4 & 50 & 23 $\pm$ 13 \\
Serpens\tablenotemark{b}   &  0.0 & 1.2 & 20 & 12 $\pm$ 3  \\
\enddata
\tablecomments{Median difference between the different ways to estimate
               extinction within a molecular cloud (IRIS $A_V$ - 
	       NIR/optical $A_V$), and the scatter between the methods 
	       (see Section \ref{TEMPEXT}).}
\tablenotetext{a}{IRAS flux normalized to the 2MASS/NICER extinction}
\tablenotetext{b}{IRAS flux normalized to the Cambr{\'e}sy extinction}
\end{deluxetable}

\end{document}